\documentclass[aps,prb,reprint,showkeys,twocolumn,nofootinbib,superscriptaddress]{revtex4-2}
\bibpunct{[}{]}{,}{n}{}{}
\usepackage{natbib}
\usepackage{amssymb}
\usepackage{booktabs}
\usepackage{color}
\usepackage{amsfonts}
\usepackage{textcomp}
\newcommand{\ped}[1]{\ensuremath{_{\rm #1}}}
\newcommand{\apex}[1]{\ensuremath{^{\rm #1}}}

\newcommand{\mylargewidth}{0.8\textwidth}
\newcommand{\mywidth}{\columnwidth}
\newcommand{\mysmallwidth}{0.8\columnwidth}
\definecolor{blue}{rgb}{0,0,0}
\definecolor{link}{RGB}{57,106,177}
\usepackage{float}
\usepackage[caption = false]{subfig}
\usepackage{graphicx}
\usepackage{hyperref}
\hypersetup{
    colorlinks=true,
    linkcolor=link,
    citecolor=link,
    urlcolor=link
}
\AtBeginDocument{
\heavyrulewidth=.08em
\lightrulewidth=.05em
\cmidrulewidth=.03em
\belowrulesep=.65ex
\belowbottomsep=0pt
\aboverulesep=.4ex
\abovetopsep=0pt
\cmidrulesep=\doublerulesep
\cmidrulekern=.5em
\defaultaddspace=.5em
}
\begin{document}

\title{Orientation-dependent electric transport and band filling in hole co-doped epitaxial diamond films}

\author{Erik Piatti}
\email{erik.piatti@polito.it}
%\author{Davide Romanin}
\affiliation{\mbox{Department of Applied Science and Technology, Politecnico di Torino, I-10129 Torino, Italy}}
\author{Alberto Pasquarelli}
\affiliation{\mbox{Institute of Electron Devices and Circuits, Ulm University, 89069 Ulm, Germany}}
\author{Renato S. Gonnelli}
\email{renato.gonnelli@polito.it}
\affiliation{\mbox{Department of Applied Science and Technology, Politecnico di Torino, I-10129 Torino, Italy}}

\begin{abstract}
Diamond, a well-known wide-bandgap insulator, becomes a low-temperature superconductor upon substitutional doping of carbon with boron. However, limited boron solubility and significant lattice disorder introduced by boron doping prevent attaining the theoretically-predicted high-temperature superconductivity. Here we present an alternative co-doping approach, based on the combination of ionic gating and boron substitution, in hydrogenated thin films epitaxially grown on (111)- and (110)-oriented single crystals. Gate-dependent electric transport measurements show that the effect of boron doping strongly depends on the crystal orientation. In the (111) surface, it strongly suppresses the charge-carrier mobility and moderately increases the gate-induced doping, while in the (110) surface it strongly increases the gate-induced doping with a moderate reduction in mobility. In both cases the maximum total carrier density remains below $2{\cdot}10^{14}$~cm\apex{-2}, three times lower than the value theoretically required for high-temperature superconductivity. Density-functional theory calculations show that this strongly orientation-dependent effect is due to the specific energy-dependence of the density of states in the two surfaces. Our results allow to determine the band filling and doping-dependence of the hole scattering lifetime in the two surfaces, showing the occurrence of a frustrated insulator-to-metal transition in the (110) surface and of a re-entrant insulator-to-metal transition in the (111) surface.

\bigskip
\textbf{Cite this article as:} 

E. Piatti, A. Pasquarelli, and R. S. Gonnelli. \textit{Appl. Surf. Sci.} \textbf{528}, 146795 (2020).

\bigskip
\textbf{DOI:} 

\href{https://doi.org/10.1016/j.apsusc.2020.146795}{10.1016/j.apsusc.2020.146795}
\end{abstract}


\maketitle

\section{Introduction}\label{sec:introduction} 

Diamond, a band insulator with a wide band gap $\sim 5.5$~eV, holds a great potential for both fundamental and applied science, thanks to its large thermal conductivity and intrinsic charge-carrier mobility, excellent electrochemical stability, high breakdown electric field, and biocompatibility~\cite{PanBook}. Two main methods exist to induce hole-type conductivity in diamond: boron (B) substitution to carbon (C) atoms in the bulk lattice, and hydrogen (H) termination of the dangling bonds at the surface. Since B dopants remove electrons from the valence band of diamond, B substitution at very large concentrations ($> 10^{20}$~cm\apex{-3}) is able to induce a full insulator-to-metal transition and superconductivity (SC) at low temperature \cite{EkimovNature2004, BustarretPRL2004, YokoyaNature2005}. At the same time, however, it introduces significant lattice disorder, degrading its otherwise excellent normal-state and SC transport properties \cite{TakanoAPL2004, IshizakaPRL2007, BustarretPSSA2008, OkazakiAPL2015}. Indeed, disorder and limited B solubility have so far been the major obstacle in exploiting its very large Debye temperature ($\sim2000$~K) to experimentally attain its predicted high-temperature superconductivity~\cite{BoeriPRL2004, LeePRL2004, BoeriJPCS2006, GiustinoPRL2007}. Alternatively, when H-terminated diamond surfaces are exposed to electron-accepting molecules (such as air moisture), electron transfer can occur from the valence band of diamond to the empty energy levels of the adsorbates \cite{MaierPRL2000, StrobelNature2004}. This results in the formation of two-dimensional hole gases (2DHGs) with a low hole density ($\lesssim 10^{13}$~cm\apex{-2}) that can be further tuned by a transverse electric field~\cite{LandstrassAPL1989, MaierPRL2000, StrobelNature2004, EdmondsNanoLett2015}. Notably, these 2DHGs are bound to the surface since the field-induced hole density extends by less than 1~nm from the first C layer for typical intensities of the transverse electric field~\cite{NakamuraPRB2013, SanoPRB2017, RomaninApSuSc2019, RomaninApSuSc2020}, although their confinement may be slighly reduced in the presence of ultrahigh electric fields which are able to penetrate further into the bulk~\cite{FeteAPL2016, PiattiPRB2017, ValentinisPRB2017, PiattiApSuSc2018nbn, UmmarinoPRB2017}. These 2DHGs have also been predicted to host electric-field-induced superconductivity, when the induced surface hole densities become large enough~\cite{NakamuraPRB2013, SanoPRB2017, RomaninApSuSc2019, RomaninApSuSc2020}. However, the extensive experimental studies carried out on the electric transport properties of gated H-terminated diamond surfaces have not yet reported evidence for either superconductivity or good metallic behavior, likely due to the low densities ($\lesssim 7\cdot10^{13}$~cm\apex{-2}) that were achieved~\cite{PiattiLTP2019, YamaguchiJPSJ2013, TakahidePRB2014, AkhgarNanoLett2016, TakahidePRB2016, AkhgarPRB2019}. This was relatively surprising, since these studies employed the ionic gating technique, which can potentially tune surface carrier densities of the order of $10^{15}$~cm\apex{-2}~\cite{PiattiPRB2017, DagheroPRL2012, LiNature2016, XiPRL2016} thanks to the extremely large geometrical capacitance of the electric double layer (EDL) at an electrolyte/electrode interface~\cite{FujimotoPCCP2013, UenoJPSJ2014}. However, we recently demonstrated~\cite{PiattiEPJ2019} that the presence of a finite B doping allowed to increase the maximum sheet carrier density tunable in EDL-transistors (EDLTs) realized on nanocrystalline diamond films by a factor $\sim 3$, an approach that could potentially be transferable to high-mobility epitaxial films.

In this work, we build upon our previous results and perform a doping-dependent characterization of the electric transport properties of H-terminated diamond thin films epitaxially grown on oriented single-crystal substrates by means of a co-doping approach, were ionic gating and B substitution are simultaneously employed to tune the free hole density in a thin surface layer. We focus our attention to (111)- and (110)-oriented facets for thin film growth, since these are the crystal orientations where field-induced superconductivity has been predicted to emerge at sufficently large values of hole density~\cite{NakamuraPRB2013, SanoPRB2017, RomaninApSuSc2019, RomaninApSuSc2020}. While the values of hole density theoretically required for high-temperature superconductivity ($\sim 6\cdot10^{14}$~h\apex{+}cm\apex{-2} \cite{RomaninApSuSc2019, RomaninApSuSc2020, PiattiLTP2019}) remain outside of our reach, we find evidence of a strong sensitivity to the crystal orientation in the behavior of the ion-gated epitaxial films upon the introduction of B dopants. We find that B substitution introduces a slightly larger free hole density in the (111)-oriented films than in the (110)-oriented ones, at the expense of a much larger suppression in the hole mobility. Most strikingly, we also discover that the B doping leads to a five-fold increase in the gate capacitance in the (110)-oriented films, while leaves that of the (111)-oriented films almost unaffected. By employing \textit{ab initio} calculations of the electronic bandstructure of the doped surfaces, we are able to directly link this orientation-dependent capacitance enhancement to the specific energy-dependence of the electronic density of states in the two surfaces, which leads to starkly different doping-dependencies of their quantum capacitance. By combining our gate-dependent transport measurements with the calculations, we also estimate the doping-dependencies of the hole scattering lifetime in the two surfaces, which reveal that the combination of ionic gating and B doping tunes the (110) and (111) surfaces across a frustrated and re-entrant insulator-to-metal transition respectively.

\section{Results}

\begin{figure*}
\begin{center}
\includegraphics[keepaspectratio, width=\textwidth]{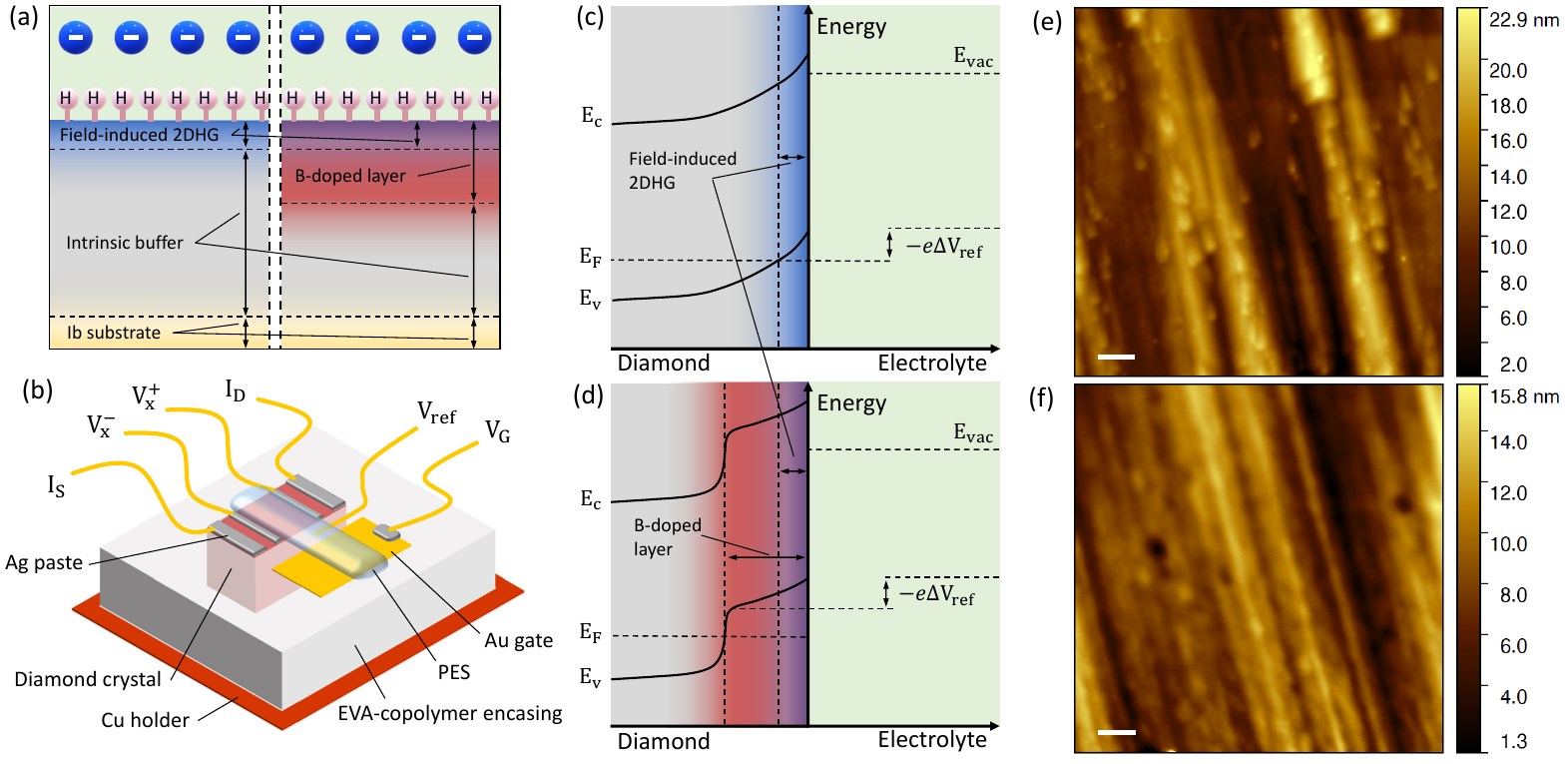}
\end{center}
\caption {
(a) Schematic structure of the ion-gated, H-terminated (left) and B-doped (right) diamond epitaxial films. False-colors highlight the different components of the sample: Single-crystal Ib diamond substrate (yellow), intrinsic buffer layer (grey), B-doped $\delta$-layer (red, right only), and electric-field-induced 2DHG (blue). Pink spheres represent the H termination, also present in the B-doped samples. Blue spheres represent the TFSI\apex{-} anions in the EDL. 
(b) Sketch of a complete diamond-based EDLT, including electrical contacts for four-wire resistance measurements, as well as gate and reference electrodes.{\color{blue}
(c) Pictorial depiction of the expected band diagram of an ion-gated, H-terminated epitaxial diamond film~\cite{DankerlPRL2011}. Conduction band minimum $E\ped{c}$, valence band maximum $E\ped{v}$, Fermi level $E\ped{F}$, vacuum level $E\ped{vac}$, and gate-induced shift of the reference potential $\Delta V\ped{ref}$ are highlighted. The color coding follows that in panel (a).
(d) Same as (c) for a B-doped epitaxial film.}
(e) AFM topography map of the surface of a diamond film grown epitaxially on a (111)-oriented substrate. 
(f) Same as (c) for a film grown on a (110)-oriented substrate. Scale bars are $1\mu$m.
} \label{figure:device}
\end{figure*}

\subsection{Device fabrication}\label{sec:fabrication}

\begin{table}[b]
\begin{tabular*}{\mywidth}{r @{\extracolsep{\fill}} cllrr@{}}
\toprule
\multicolumn{3}{c}{Operation}	& Chemical	& Temperature	& Time \\ \midrule
1	& & Sonication	& Acetone	& 25\apex{\circ}C	& 10 min	\\
2	& & Sonication	& Isopropanol	& 25\apex{\circ}C	& 10 min	\\
3	& & Etching	& Chromosulphuric acid	& 100\apex{\circ}C	& 20 min	\\
4	& & Rinsing		& DI-water flow	& 25\apex{\circ}C	& 15 min	\\
5	& & Etching		& HCl:H\ped{2}O\ped{2} = 1:1	& 80\apex{\circ}C	& 10 min	\\
6	& & Rinsing		& DI-water	& 25\apex{\circ}C	& $\lesssim$~5 min	\\
7	& & Etching		& H\ped{2}O\ped{2}:NH\ped{4}OH = 1:1	& 120\apex{\circ}C	& 15 min	\\
8	& & Rinsing		& DI-water	& 25\apex{\circ}C	& $\lesssim$~5 min	\\
9	& & Etching		& HCl:HNO\ped{3} = 3:1	& 25\apex{\circ}C	& overnight	\\
10	& & Rinsing		& DI-water	& 25\apex{\circ}C	& $\lesssim$~5 min	\\
11	& & Etching		& H\ped{2}SO\ped{4}:H\ped{2}O\ped{2} = 2:1	& 120\apex{\circ}C	& 10 min	\\
12	& & Rinsing		& DI-water flow	& 25\apex{\circ}C	& 20 min	\\
\bottomrule
\end{tabular*}
\caption{Substrate cleaning procedure performed before thin-film growth. After steps 2 and 12, the substrates were blow-dried with a nitrogen gun.}
\label{tab:substrate_cleaning}
\end{table}

We selected commercial (111)- and (110)-oriented Ib-type single crystals grown by the high-temperature high-pressure method (Sumitomo) as substrates for thin-film growth. Prior to the growth process, the substrates were thoroughly cleaned by the procedure detailed in Table~\ref{tab:substrate_cleaning}. A nominally undoped, 100 nm-thick homoepitaxial buffer layer was first grown to avoid any influence from the nitrogen dopants in the substrates. An additional 2 nm-thick B-doped $\delta$-layer was subsequently grown on half the samples. Growths were performed in two specifically process-dedicated reactors -- in order to avoid any B contamination of the intrinsic layer -- in H\ped{2}/CH\ped{4} atmosphere, resulting in H-termination of all the surfaces. The B dopants were supplied by a solid B-coated wire (Goodfellow) inserted into the MPCVD chamber by means of a magnetic manipulator. {\color{blue}Growths were carried out at a pressure of 2~kPa, a reactor temperature of 750\apex{\circ}C, a plasma RF power of 750~W and an H\ped{2} flow of 200 sccm and varying CH\ped{4} concentration. In particular, the $\delta$-layers were grown by inserting the boron rod in pure H\ped{2} plasma for 2 s, followed by a 5 s growth step under a dilute H\ped{2}/CH\ped{4} pulse, and subsequently switching off the plasma and evacuating the gas mixture. Growths were performed on all samples simultaneously to minimize sample-to-sample variations in the $\delta$-layer thickness. Nevertheless, since the growth conditions were chosen} in order to match the calibrated growth rate on (100)-oriented single crystals~\cite{ElHajjDRM2008, ScharpfPSSA2013}, the reported thicknesses should be treated as nominal values only. Consequently, in the following we will limit our discussion to quantities (carrier density, conductivity, mobility, capacitance) per unit surface, which remain well-defined even in the absence of a well-defined sample thickness. The resulting diamond layer stack is sketched in Fig.~\ref{figure:device}(a) for the two types of films (H-terminated and B-doped){\color{blue}, and a pictorial view of the expected band diagrams of the gated films -- based on the band diagram reported in Ref.~\onlinecite{DankerlPRL2011} -- is shown in Fig.~\ref{figure:device}(c,d).}

After the growth, all substrate facets except the top one were etched and oxidized via argon/oxygen plasma, leaving only the top facet conductive. The surface morphology of the films was then inspected via Atomic Force Microscopy (AFM). In Fig.~\ref{figure:device}(e,f) we show two 10$\mu$m$\times$10$\mu$m topography maps acquired with a Bruker Innova microscope in tapping mode on a (111) and (110)-oriented film respectively. The most prominent features correspond to smooth, oriented trenches extending across both surfaces, which we ascribe to the substrate polishing process and are replicated at the film surface due to the conformal growth. We do not observe the presence of rectangular or triangular grains in the topography, thus excluding a significant degree of polycrystalline growth \cite{PiattiEPJ2019,UshizawaDRM1998}. We do, however, observe the presence of sharp ``cusps" scattered throughout the (111) surface. {\color{blue}Having excluded the possibility of these features being due to tip imaging artifacts by obtaining consistent results over different surfaces and using multiple tips, these} could either be due to an imperfect cleavage of the substrate along its crystalline axis, or to an imperfect conformal growth of (111)-oriented facets in the film~\cite{UshizawaDRM1998}. The mean square roughness of the two surfaces ($S\ped{q}|\ped{(111)}\simeq 3.8\,\mathrm{nm}$ and $S\ped{q}|\ped{(110)}\simeq 2.4\,\mathrm{nm}$) is typical of polished Ib crystals \cite{ElHajjDRM2008} and one order of magnitude smaller than those of nanocrystalline films \cite{PiattiEPJ2019}. On the other hand, the peak-to-peak roughness  of our samples is $\sim23$ and $\sim16$~nm respectively: Assuming a nominal $\delta$-layer thickness of 2~nm, we cannot in principle rule out the presence of discontinuities in the conducting surface layer, leading to a degradation in the macroscopic conductivity and mobility, as well as to an increase in the degree of disorder. This issue could be addressed in future works by optimizing the growth of the buffer layer, which has been shown to allow reducing $S\ped{q}$ down to the sub-nm range~\cite{ElHajjDRM2008}.

After this preliminary characterization, the epitaxial films were incorporated into the EDLT architecture. As sketched in Fig.~\ref{figure:device}(b), the few-mm thick diamond substrates were encased in a plastic holder, which was sealed by briefly heating the assembly above the melting point of the plastic ($\sim 70^{\circ}$C). Electrical contacts for four-wire sheet conductivity measurements (source, drain and longitudinal voltage probes) were realized by drop-casting a small amount of silver conductive paste. The side gate was realized by glueing a thin gold leaf to the plastic holder at the same level of the diamond surface. Additionally, a thin gold wire was fixed in close proximity to both the gate and the sample to act as the reference electrode. Finally, the liquid precursor to the polymer-electrolyte system (PES) was drop-casted on the side gate, the reference electrode and the diamond surface between the voltage contacts and UV-cured in the controlled atmosphere of a dry room, resulting in a rubbery, mechanically-stable ion-gel. Following our earlier experiments on nanocrystalline diamond films \cite{PiattiEPJ2019}, the PES was composed of a misture of Bisphenol A ethoxylate dimethacrylate oligomer (BEMA) and 1-Ethyl-3-methylimidazolium bis(trifluoromethylsulfonyl)imide ionic liquid (EMIM-TFSI) in a 7:3 weight ratio, together with 3wt.\% of Darocur 1173 photoinitiator. The complete device was then mounted on the cold finger of a Cryomech pulse-tube cryocooler and left to degas in high vacuum ($p\lesssim 10^{-5}$ mbar) overnight to minimize water traces absorbed in the PES.

\subsection{Ionic-gate operation}\label{sec:ionic_gate}
%\begin{figure}
%\begin{center}
%%\includegraphics[keepaspectratio, width=0.75\textwidth]{Figure_EDLT.eps}
%\end{center}
%\caption {
%(a) 
%(b) 
%(c) 
%(d) 
%} \label{figure:EDLT}
%\end{figure}

\begin{figure*}
\begin{center}
\includegraphics[keepaspectratio, width=\mylargewidth]{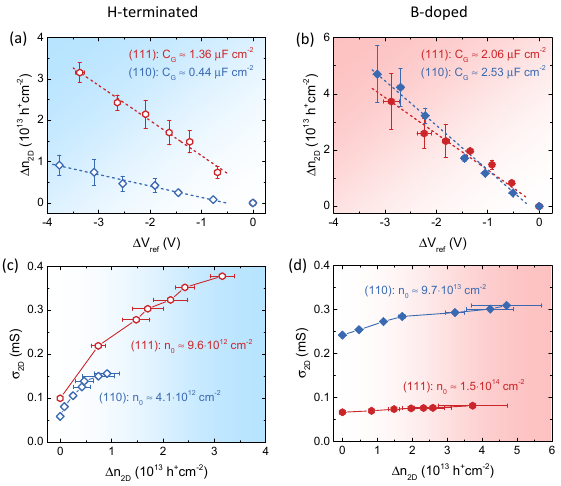}
\end{center}
\caption {
(a) Gate-induced hole density per unit surface as a function of the gate-induced voltage drop across the electrolyte/diamond interface for H-terminated (111)-oriented (red hexagons) and (110)-oriented (blue diamonds) films.
(b) Same as (a) for B-doped films. Dashed lines are the linear fits to the data from which the gate capacitances reported in the legend are estimated.
(c) Sheet conductance as a function of the gate-induced hole density for H-terminated (111)-oriented (red hexagons) and (110)-oriented (blue diamonds) films.
(d) Same as (c) for B-doped films. Legend shows also the hole density at $V\ped{G}=0$ estimated by linear regression (see the main text for details).
} \label{figure:gating_operation}
\end{figure*}

Gate-dependent electric transport measurements were performed at the optimal operation temperature \mbox{$T\simeq 240$~K}, slightly above the glass transition temperature of the PES, to minimize the chance of activating electrochemical reactions at the H-terminated surface. The sheet resistance $R\ped{s}$ of the samples was measured by applying a small DC current $I\ped{DS}\sim 1\mu$A between the drain and source contacts and measuring the longitudinal voltage drop $V\ped{xx}$ across the voltage contacts. Common mode offsets such as thermoelectric voltages and (small) contributions from the gate current were removed by measuring $R\ped{s}$ with both signs of $I\ped{DS}$ and taking the average \cite{DagheroPRL2012}. The sheet conductance was determined as $\sigma\ped{2D}=1/R\ped{s}$. Gate voltages $V\ped{G}$ between $-1$~V and $-6$~V were applied between the gate and source contacts while simultaneously measuring the gate current $I\ped{G}$ and the voltage drop $V\ped{ref}$ between the reference electrode and the source. $I\ped{DS}$, $V\ped{G}$ and $I\ped{G}$ were applied and measured with the two independent channels of an Agilent B2912 source-measure unit. $V\ped{xx}$ and $V\ped{ref}$ were measured with an Agilent 34420 nanovoltmeter and a Keithley 2182 nanovoltmeter respectively. {\color{blue}The values of applied $V\ped{G}$ were limited to $-6$~V due to the finite electrochemical stability window of the electrolyte, as detected by a large increase in $I\ped{G}$ whenever negative values of $V\ped{G}$ in excess of $-6$~V were applied across the electrolyte.}

Each gate-dependent measurement was performed according to the procedure detailed in Refs.~\onlinecite{DagheroPRL2012, TortelloApsusc2013, PiattiJSNM2016, PiattiPRM2019, PiattiEPJ2019}. Briefly, the selected value of $V\ped{G}$ was applied and removed in a step-like fashion, and the resulting transients on $\sigma\ped{2D}$, $I\ped{G}$ and $V\ped{ref}$ were acquired. After each step-like application/removal, the ion dynamics were allowed to settle for at least $\sim10$ min. Each $V\ped{G}$ value was applied and removed multiple times to ensure the reproducibility of the measurements. Double-step chroconoulometry (DSCC), a well-established electrochemical technique \cite{ScholtzBook} that allows to determine the amount of charge stored in the EDL during the charging/discharging of an EDLT \cite{DagheroPRL2012, TortelloApsusc2013, PiattiJSNM2016, PiattiPRM2019, PiattiEPJ2019}, was employed to determine the gate-induced hole density $\Delta n\ped{2D}$ at the diamond surface from the $I\ped{G}$ transients. The gate-induced voltage drop across the electrolyte/diamond interface was directly monitored as $\Delta V\ped{ref} = V\ped{ref}(V\ped{G}) - V\ped{ref}(0)$: This removes parasitic voltage drops across the gate/electrolyte interface, and is mandatory for a reliable estimation of the gate capacitance \cite{ZhangNanoLett2019}. 

We first discuss the dependencies of $\Delta n\ped{2D}$ on $\Delta V\ped{ref}$ in our ion-gated diamond films as shown in Fig.~\ref{figure:gating_operation}(a,b). For simplicity, in the following we shall refer to H-(111) and H-(110) to the H-terminated films and B-(111) and B-(110) to the B-doped films. In all cases, the sheet carrier density increases with increasingly negative $\Delta V\ped{ref}$ with a nearly linear scaling, allowing for a reliable estimation of the gate capacitance $C\ped{G}$ from the slope of the linear fit. All values $C\ped{G}$ will be given with an uncertainty of $30\%$, which is the typical uncertainty of $\Delta n\ped{2D}$ measured by DSCC. In the H-terminated surfaces [see Fig.~\ref{figure:gating_operation}(a)] $C\ped{G}$ and the maximum achieved $\Delta n\ped{2D}$ are strongly orientation-dependent. In the H-(111) surface, we find $C\ped{G}=1.36\pm0.41\,\mu\mathrm{F\,cm^{-2}}$, comparable to the values reported in the literature ($2.6-4.6\,\mu\mathrm{F\,cm^{-2}}$)~\cite{YamaguchiJPSJ2013, TakahidePRB2014, PiattiLTP2019}, whereas in the H-(110) surface we find a much smaller value $C\ped{G}=0.44\pm0.13\,\mu\mathrm{F\,cm^{-2}}$. As a consequence, we find that the maximum achieved $\Delta n\ped{2D}$ in the H-(111) surface is $\Delta n\ped{2D}\apex{max} = 3.2\pm0.3\cdot 10^{13}$~h\apex{+}cm\apex{-2}, again comparable with the literature values~\cite{YamaguchiJPSJ2013, TakahidePRB2014, PiattiLTP2019}, and much larger than the value in the H-(110) surface $\Delta n\ped{2D}\apex{max} = 9.1\pm2.4\cdot 10^{12}$~h\apex{+}cm\apex{-2}. Conversely, in the B-doped surfaces [see Fig.~\ref{figure:gating_operation}(b)] both $C\ped{G}$ and $\Delta n\ped{2D}\apex{max}$ are found to be slighly larger in the (110) surface, although the differences are much less pronounced than in the H-terminated surfaces: We find $C\ped{G}=2.06\pm0.62\,\mu\mathrm{F\,cm^{-2}}$ and $\Delta n\ped{2D}\apex{max} = 3.7\pm1.0\cdot 10^{13}$~h\apex{+}cm\apex{-2} for the B-(111) surface and $C\ped{G}=2.53\pm0.76\,\mu\mathrm{F\,cm^{-2}}$ and $\Delta n\ped{2D}\apex{max} = 4.7\pm1.0\cdot 10^{13}$~h\apex{+}cm\apex{-2} for the B-(110) surface. Overall, we find that the B-doping process leads to more than a five-fold enhancement in $C\ped{G}$ and $\Delta n\ped{2D}\apex{max}$ in the (110) surface, while it leaves those in the (111) surface almost unaffected. Notably, even in the B-doped epitaxial films $C\ped{G}$ and $\Delta n\ped{2D}\apex{max}$ reach up only to about one-third of their values in thick (300~nm) B-doped nanocrystalline diamond films~\cite{PiattiEPJ2019, PiattiLTP2019}, suggesting that the thickness of the B-doped diamond layer may play an important role in the enhancement of the gate capacitance.

We now consider the dependencies of $\sigma\ped{2D}$ on $\Delta n\ped{2D}$ in our ion-gated diamond films as shown in Fig.~\ref{figure:gating_operation}(c,d). In all films, $\sigma\ped{2D}$ increased monotonically at the increase of the induced hole density, showing signs of incipient saturation at large $\Delta n\ped{2D}$. Notably, we did not observe significant decreases of $\sigma\ped{2D}$ over time when large negative $\Delta V\ped{ref}$ were being held at the electrolyte/diamond interface. This suggests that the interface between hydrogenated diamond and EMIM-TFSI ionic liquid was stable when operated at $T\simeq 240$~K, and electrochemical reactions -- such as the anodic oxidation of the H-terminated surface, which was reported to possibly become an issue~\cite{YamaguchiJPSJ2013} -- were mostly negligible. In all of our films, a finite $\sigma\ped{2D}$ was observed also at $V\ped{G}=0$, indicating the presence of a finite `` intrinsic" density of free holes, $n_0$: In the B-doped surfaces, this can be directly ascribed to the presence of the B dopants. In the H-terminated surfaces, it indicates that electron-accepting molecules were adsorbed to the diamond surface, resulting in a finite free hole density from charge-transfer~\cite{MaierPRL2000, StrobelNature2004}. In turn, this suggests that either residual surface adsorbates due to air exposure of the film became trapped during PES drop-casting, or that the polymeric chains in the PES itself acted as electron acceptors. Thanks to the monotonic increase of $\sigma\ped{2D}$ with $\Delta n\ped{2D}$ in all our films, the value of $n_0$ for the different surfaces can be estimated by linear regression of the experimental data to $\sigma\ped{2D} = 0$. This gives $n_0 = 9.6\pm 2.9 \cdot 10^{12}$~h\apex{+}cm\apex{-2} for the H-(111) surface, $n_0 = 4.1\pm 1.2 \cdot 10^{12}$~h\apex{+}cm\apex{-2} for the H-(110) surface, $n_0 = 1.5\pm 0.4 \cdot 10^{14}$~h\apex{+}cm\apex{-2} for the B-(111) surface, and $n_0 = 9.7\pm 2.9 \cdot 10^{13}$~h\apex{+}cm\apex{-2} for the B-(110) surface. As expected, the values of $n_0$ introduced by the B-doping process are more than one order of magnitude larger than those associated with H-assisted charge transfer, and the (111) surface shows larger values of $n_0$ than the (110) surface both in H-terminated and B-doped films. The large difference in $n_0$ between H-terminated and B-doped films also accounts for their very different tunability upon application of a finite $V\ped{G}$: $\sigma\ped{2D}$ in the H-terminated films (where $n_0 \ll \Delta n\ped{2D}\apex{max}$) can be tuned by $\sim 300\%$ of its value at $V\ped{G}=0$, whereas in the B-doped films (where $n_0 \gg \Delta n\ped{2D}\apex{max}$) only by $\sim20\%$.

\begin{figure}
\begin{center}
\includegraphics[keepaspectratio, width=\mysmallwidth]{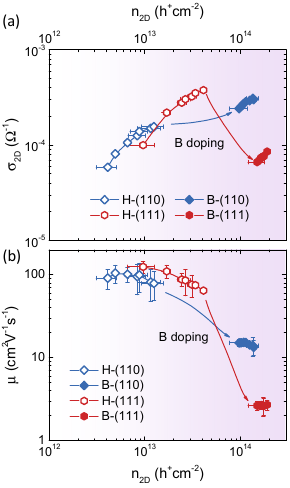}
\end{center}
\caption {
Sheet conductance (a) and hole mobility (b) as a function of the total carrier density per unit surface in the ion-gated epitaxial diamond films. Hollow (filled) red hexagons refer to the H-terminated (B-doped) (111) films, hollow (filled) blue diamonds to the H terminated (B-doped) (110) films.
} \label{figure:mobility}
\end{figure}

Having determined both the intrinsic and gate-induced hole densities, $n_0$ and $\Delta n\ped{2D}$, we can also recover the total hole density in our ion-gated films as $n\ped{2D} = n_0 + \Delta n\ped{2D}$. In Fig.~\ref{figure:mobility}(a) we map the  dependence of $\sigma\ped{2D}$ in all ion-gated films as a function of the \textit{total} carrier density per unit surface. The H-terminated surfaces show a very similar dependence of $\sigma\ped{2D}$ on $n\ped{2D}$, suggesting that the two surfaces share comparable values of hole mobility. The main difference between the two lies in the fact that the H-(111) surface is able to accomodate a larger amount of field-induced free carriers and can thus become more conducting than the H-(110) surface. The behavior of the B-doped surfaces is more complex. In both B-(111) and B-(110) $\sigma\ped{2D}$ increases with increasing $n\ped{2D}$; however, the introduction of B dopants \textit{suppresses} $\sigma\ped{2D}$ in the B-(111) surface below its minimum value in the H-(111) surface, whereas it further \textit{enhances} $\sigma\ped{2D}$ in the B-(110) surface above its maximum value in the H-(110) surface. This indicates that the B-doping process affects in a very different way the mobility in the two surfaces.

We assess this quantitatively in Fig.~\ref{figure:mobility}(b), where we plot the $n\ped{2D}$-dependence of the hole mobility \mbox{$\mu = \sigma\ped{2D}/en\ped{2D}$} where $e$ is the elementary charge. As expected, the H-(111) and H-(110) surface shows very similar values of $\mu$ and similar $n\ped{2D}$-dependencies: As already reported in the case of ion-gated H-(111) and H-(100) surfaces at $T > 200$~K~\cite{YamaguchiJPSJ2013, PiattiLTP2019}, $\mu\sim 100$~cm\apex{2}V\apex{-1}s\apex{-1} near $V\ped{G} = 0$ and then rapidly decreases at the increase of $n\ped{2D}$. This behavior has been explained as due to a combination of the intrinsic doping dependence of acoustic-phonon scattering~\cite{YamaguchiJPSJ2013, PiattiLTP2019, RezekAPL2006, KasuAPE2012} and the gate-induced disorder caused by the ions in the EDL acting as charged scattering centers~\cite{PiattiEPJ2019, PiattiLTP2019, GallagherNatCommun2015, SaitoACSNano2015, Gonnelli2dMater2017, PiattiApSuSc2017, PiattiNanoLett2018, PiattiApSuSc2018mos2, OvchinnikovNatCommun2016, PiattiAPL2017, LuPNAS2018, PiattiPRM2019}. The introduction of B dopants strongly suppresses $\mu$ in both diamond surfaces, but with a starkly different effectiveness. In the B-(110) surface, $\mu\sim15$~cm\apex{2}V\apex{-1}s\apex{-1} and is still weakly decreasing at the increase of $n\ped{2D}$. Moreover, the reduction in $\mu$ between the H-(110) and B-(110) surface is ``smooth" as it does not deviate too much from a simple extrapolation of the $n\ped{2D}$-dependence in the H-(110) surface. In the B-(111) surface, on the other hand, $\mu\sim 3$~cm\apex{2}V\apex{-1}s\apex{-1} is nearly constant with increasing $n\ped{2D}$ and deviates from a simple extrapolation of the $n\ped{2D}$-dependence in the H-(111) surface by about one order of magnitude. These findings suggest that the B dopants suppress $\mu$ in the (110) surface mainly due to its intrinsic doping dependence, with a minimal contribution coming from the introduction of extrinsic disorder, whereas this latter contribution appears to be dominant in the (111) surface. We summarize the main physical parameters obtained by the gate-dependent transport measurements in Table~\ref{tab:electric_transport}.

\begin{table}[]
\begin{tabular*}{\mywidth}{c @{\extracolsep{\fill}} ccccc@{}}
\toprule
        & $C\ped{G}$                & $n_0$            & $\Delta n\ped{2D}\apex{max}$    & $\mu\apex{min} \div \mu\apex{max}$ \\ \midrule
        & $\mathrm{\mu F\,cm^{-2}}$ & \multicolumn{2}{c}{$10^{13}\,\mathrm{h^+cm^{-2}}$} & $\mathrm{cm^2V^{-1}s^{-1}}$        \\ \midrule
H-(111) & $1.36\pm0.41$             & $0.96\pm0.29$    & $3.2\pm0.3$                     & $53 \div 145$                      \\
H-(110) & $0.44\pm0.13$             & $0.41\pm0.12$    & $0.91\pm2.4$                    & $47 \div 138$                      \\
B-(111) & $2.06\pm0.62$             & $15\pm4$         & $3.7\pm1.0$                     & $1.9 \div 3.3$                     \\
B-(110) & $2.53\pm0.76$             & $9.7\pm2.9$      & $4.7\pm1.0$                     & $10 \div 18$                       \\ \bottomrule
\end{tabular*}
\caption{Gate capacitance, intrinsic hole density, maximum gate-induced hole density, and mobility range (including the experimental uncertainty) in all ion-gated diamond surfaces.}
\label{tab:electric_transport}
\end{table}

\subsection{Electronic bandstructure}\label{sec:bandstructure}

To obtain further insight in the orientation-dependent response to B doping of our ion-gated diamond films, we carried out \textit{ab initio} density functional theory (DFT) calculations as implemented in the Quantum ESPRESSO package~\cite{QEspresso} for the electronic bandstructure and the density of states (DOS) in H-terminated, hole-doped (111) and (110) diamond surfaces. Exchange-correlations corrections were included via the Perdew-Burke-Ernzerhof (PBE) exchange-correlation functional~\cite{PerdewPRL1996}. Core-electrons contributions of C atoms were modelled with a projector-augmented wave pseudopotential (PAW~\cite{BlochlPRB1994}) while those of H atoms with a norm-conserving Vanderbilt pseudopotential~\cite{HamannPRB2013}, both
obtained from the standard solid-state pseudopotential (SSSP) precision library~\cite{SSSP}. Energy cut-offs for the valence-electron wavefunction and for the density were set to 65 Ry and 600 Ry respectively. Convergence conditions for the self-consistent solution of the Kohn-Sham problem were set to 10\apex{-9} Ry for the total energy and to 10\apex{-3} Ry/Bohr for the total force acting on the atoms. Brillouin zone integrations were performed on Monkhorst-Pack grids~\cite{MonkhorstPRB1976} composed of $24\times24\times1$ \textbf{k}-points for the (111) surface and $24\times16\times1$ \textbf{k}-points for the (110) surface, using a Gaussian smearing of 6~mRy. For each self-consistent calculation, a preliminary relaxation of the atomic positions was carried out. The DOS were calculated on Monkhorst-Pack grids using twice the number of \textbf{k}-points and tetrahedra occupations to ensure better convergence of the results.

\begin{figure*}
\begin{center}
\includegraphics[keepaspectratio, width=\mylargewidth]{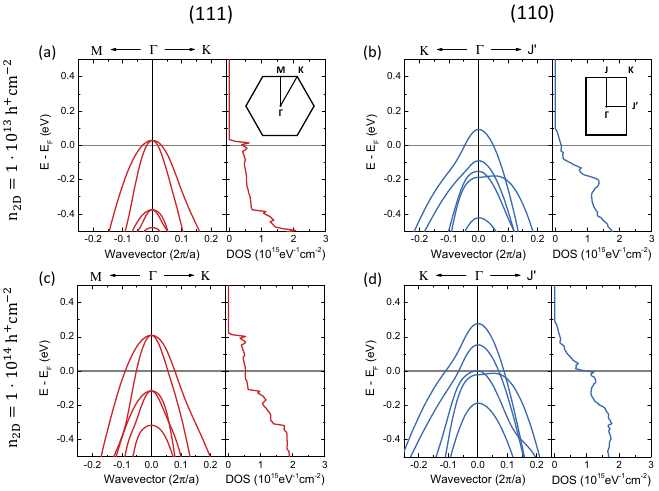}
\end{center}
\caption {
Electronic bandstructure of H-terminated diamond slabs for different crystal orientations and values of hole doping. Insets to (a,b) show the corresponding first Brillouin zones.
(a) (111)-oriented slab at $n\ped{2D}=1\cdot10^{13}$~h\apex{+}cm\apex{-2}.
(b) (110)-oriented slab at $n\ped{2D}=1\cdot10^{13}$~h\apex{+}cm\apex{-2}.
(c) (111)-oriented slab at $n\ped{2D}=1\cdot10^{14}$~h\apex{+}cm\apex{-2}.
(d) (110)-oriented slab at $n\ped{2D}=1\cdot10^{14}$~h\apex{+}cm\apex{-2}.
} \label{figure:bandstructure}
\end{figure*}

The oriented diamond surfaces were modelled following Refs.~\onlinecite{RomaninApSuSc2019, RomaninApSuSc2020}: We considered slab structures composed of 14 layers of C atoms, oriented either along the (111) or (110) directions, and terminated by a layer of H atoms on both sides. The (111)- and (110)-oriented slabs were composed by a total of 16 and 30 atoms per unit cell respectively. The in-plane lattice constant for the (111) orientation was set to $a = 2.52224$~\AA, that for the (110) orientation to $a = 3.56676$~\AA, i.e. the values computed for bulk diamond~\cite{RomaninApSuSc2019, NakamuraPRB2013}. The out-of-plane supercell was then constructed by adding $\approx30$~\AA~of vacuum between the repeated images. Charge doping was included in a Jellium model, i.e. by adding to the system a given number of holes per unit cell, together with a uniform negative background to maintain charge neutrality. Note that this approach neglects the intense electric field present at the electrolyte/diamond interface, which could be taken into account using the more computationally-demanding field-effect geometry \cite{SohierPRB2017, RomaninApSuSc2019}. However, it was explicitly demonstrated in Ref.~\onlinecite{RomaninApSuSc2019} that the bandstructure of diamond near the Fermi level showed little dependence on the electric field for $n\ped{2D}\lesssim 2\cdot10^{14}$~h\apex{+}cm\apex{-2}, which is the largest hole density we consider in this work. More importantly, in our films the hole density is only partially confined in the electric-field-induced potential well at the surface, with the largest contribution coming from the B dopants which are more appropriately approximated by the Jellium model. Since the current distribution of Quantum ESPRESSO is not able to combine the field-effect geometry with a separate Jellium doping in the same simulation, we selected the Jellium approximation as the one that more closely mimics the experimental configuration.

Overall, we performed electronic bandstructure and DOS calculations for seven different values of $n\ped{2D}$ between $3\cdot10^{12}$~h\apex{+}cm\apex{-2} and $2\cdot10^{14}$~h\apex{+}cm\apex{-2} for both the H-terminated (111)- and (110)-oriented diamond slabs. Here we will focus our discussion on the two most emblematic cases of $n\ped{2D} = 1\cdot10^{13}$ and $1\cdot10^{14}$~h\apex{+}cm\apex{-2}, as shown in Fig.~\ref{figure:bandstructure}. The bandstructures of both surfaces are in good agreement with the literature~\cite{PiattiLTP2019}: For the (111) surface [see Fig.~\ref{figure:bandstructure}(a,c)] the bandstructure is composed of two topmost valence bands, degenerate at the Brillouin-zone center and with a good parabolicity, resulting in a nearly step-like energy-dependence of the DOS; and by further lower-energy bands that remain below the Fermi level $E\ped{F}$, in good agreement with Ref.~\onlinecite{RomaninApSuSc2019}. For the (110) surface [see Fig.~\ref{figure:bandstructure}(b,d)] the energy-dependence is more complex, as was also reported in Refs.~\onlinecite{NakamuraPRB2013, SanoPRB2017}: multiple bands are present within a relatively small energy window ($\sim 300$~meV) from the top of the valence band, all of them featuring a significant degree of non-parabolic dispersion in at least one of the high-symmetry cuts across the Brillouin zone. This results in a multi-step energy dependence of the DOS, making it sensitive to hole doping.

\begin{table}[]
\begin{tabular*}{\mywidth}{c @{\extracolsep{\fill}} cccccc@{}}
\toprule
$n\ped{2D}$	& \multicolumn{2}{c}{DOS$(E\ped{F})$}	&	\multicolumn{2}{c}{N\apex{\circ} of filled bands}	& \multicolumn{2}{c}{$E\ped{v} - E\ped{F}$}	\\ \midrule
$\mathrm{h^+cm^{-2}}$	& \multicolumn{2}{c}{10\apex{14}~eV\apex{-1}~cm\apex{-2}}	&	\multicolumn{2}{c}{}	& \multicolumn{2}{c}{meV}	\\ \midrule
	& (111)	& (110)	& (111)	& (110)	& (111)	& (110)	\\ \midrule
$1\cdot10^{13}$ & 3.8	& 1.9	& 2	& 1	& 29	& 93	\\
$1\cdot10^{14}$ & 5.3	& 9.2	& 2	& 3	& 218	& 292	\\ \bottomrule
\end{tabular*}
\caption{Density of states at the Fermi level, number of filled bands, and chemical potential measured from the top of the valence band in the H-terminated (111) and (110) diamond surfaces for two different values of hole doping.}
\label{tab:bandstructure}
\end{table}

As a consequence, the band filling for increasing hole doping follows a very different behavior in the (111) and (110) surfaces. At \mbox{$n\ped{2D} = 1\cdot10^{13}$~h\apex{+}cm\apex{-2}} [see Fig.~\ref{figure:bandstructure}(a,b)], a doping level comparable to the doping range experimentally accessible in the H-terminated surfaces, two bands are filled in the (111) surface and the density of states at the Fermi level is already sizeable, DOS$(E\ped{F})\simeq3.8\cdot10^{14}$~eV\apex{-1}cm\apex{-2}. Conversely, in the (110) surface only one band is filled and the density of states is significantly smaller, DOS$(E\ped{F})\simeq1.9\cdot10^{14}$~eV\apex{-1}cm\apex{-2}. As a consequence, the chemical potential measured from the top of the valence band, $E\ped{v} - E\ped{F}$, is significantly larger in the (110) surface than the one in the (111) surface. When we consider \mbox{$n\ped{2D} = 1\cdot10^{14}$~h\apex{+}cm\apex{-2}} [see Fig.~\ref{figure:bandstructure}(c,d)], a doping level comparable to the doping range experimentally accessible in the B-doped surfaces, $E\ped{F}$ shifts downward in energy in both cases, but with quite different results. In the (111) surface, the band filling is unchanged with respect to the low-doping case, and the small non-parabolicity of the bands results only in a small increase in DOS$(E\ped{F})\simeq5.3\cdot10^{14}$~eV\apex{-1}cm\apex{-2}. In the (110) surface, on the other hand, three bands are now filled, the two at higher energy more heavily so than the third one. This strong change in the filling, together with the significant band non-parabolicity of the first two bands along the $\Gamma-$K direction, directly results in a huge increase in DOS$(E\ped{F})\simeq9.2\cdot10^{14}$~eV\apex{-1}cm\apex{-2}, which is now larger than the one in the (111) surface. The main physical parameters obtained from the bandstructures and DOS shown in Fig.~\ref{figure:bandstructure} are summarized in Table~\ref{tab:bandstructure}.

\section{Discussion}\label{sec:discussion}

\begin{figure*}
\begin{center}
\includegraphics[keepaspectratio, width=\mylargewidth]{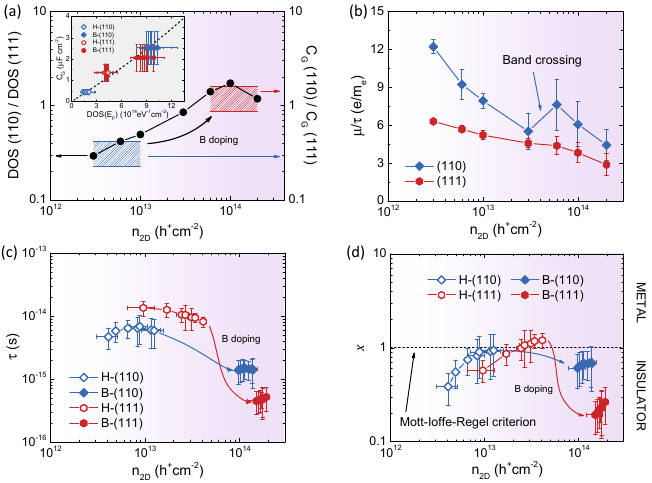}
\end{center}
\caption {
(a) Ratios of the densities of states (black dots, left scale) and gate capacitances (shaded bands, right scale) between the (110)- and (111)-oriented diamond surfaces, as a function of the total carrier density per unit surface. Inset shows the values of the gate capacitance as a function of the density of states (hollow and filled symbols), and their fit according to Eq.~\ref{eq:cap_series} (black dashed line).
(b) Mobility-to-scattering lifetime ratio as a function of the hole density for the (111)- and (110)-oriented diamond surfaces (red hexagons and blue diamonds respectively). The non-monotonicity associated to the filling of the second topmost valence band in the (110) surface is highlighted.
(c) Charge-carrier scattering lifetime, and (d) Ioffe-Regel parameter as a function of the hole density for all ion-gated epitaxial diamond films. Hollow (filled) red hexagons refer to the H-terminated (B-doped) (111) films, hollow (filled) blue diamonds to the H terminated (B-doped) (110) films. Black dashed line in (d) is the Mott-Ioffe-Regel criterion $x=1$ for the insulator-to-metal transition.
} \label{figure:discussion}
\end{figure*}

%\begin{equation}
%\frac{C\ped{G}|\ped{(110)}}{C\ped{G}|\ped{(111)}} = \frac{\mathrm{DOS}(E\ped{F})|\ped{(110)}}{\mathrm{DOS}(E\ped{F})|\ped{(111)}}
%\end{equation}

Our gate-dependent electric transport measurements and \textit{ab initio} DFT calculations show that a clear analogy exists between the behavior of the gate capacitance and that of the DOS at the Fermi level in the H-terminated (111) and (110) diamond surfaces as a function of hole doping. In the (111) surface, both quantities weakly increase upon the large increase in $n\ped{2D}$ associated to the inclusion of B dopants in the experimental samples. In the (110) surface, on the other hand, both quantities are significantly smaller for $n\ped{2D}\lesssim 1\cdot 10^{13}$~h\apex{+}cm\apex{-2}, in the absence of B dopants, and exhibit a huge increase for $n\ped{2D}\gtrsim 7\cdot 10^{13}$~h\apex{+}cm\apex{-2}, when B dopants are present in the experimental samples, eventually overcoming the corresponding values in the (111) surface. This can be quantified by plotting the doping-dependence of the ratios between the experimentally-measured values of $C\ped{G}$ in the (110)- and (111)-oriented films, together with the ratios of the theoretically-calculated values of DOS$(E\ped{F})$. As we show in Fig.~\ref{figure:discussion}(a), an extremely good match with the theoretical DOS ratio (filled black dots) is found both at low $n\ped{2D}\lesssim 1\cdot 10^{13}$~h\apex{+}cm\apex{-2} in the H-terminated surfaces (blue shaded band) and at large $n\ped{2D}\gtrsim 7\cdot 10^{13}$~h\apex{+}cm\apex{-2} in the B-doped surfaces (red shaded band), with the theoretical DOS ratio nicely reproducing the huge increase in $C\ped{G}$ upon the introduction of B dopants. This proportionality between $C\ped{G}$ and DOS$(E\ped{F})$ is typical of field-effect devices where the total gate capacitance is dominated by the quantum capacitance $C\ped{Q}=e^2\mathrm{DOS}(E\ped{F})$~\cite{LuryiAPL1988, EisensteinPRL1992, GiulianiBook2005, IlaniNatPhys2006, YuPNAS2013}. This is often observed in ion-gated devices due to the extremely large geometrical capacitance of the Helmholtz layer $C\ped{H}$, since the two contributions sum in series to determine $C\ped{G}$~\cite{YePNAS2011, GonnelliSciRep2015, PiattiApSuSc2017, Gonnelli2dMater2017, UesugiSciRep2013, ChuSciRep2014, LeNS2019, ZhangNanoLett2019}:
\begin{equation}
\frac{1}{C\ped{G}} = \frac{1}{C\ped{H}} + \frac{1}{\alpha e^2 \mathrm{DOS}(E\ped{F})}
\label{eq:cap_series}
\end{equation}
where $\alpha$ is a dimensionless proportionality constant that takes into account the ``cross-talk" in electrostatic screening between the ions in the EDL and the field-induced charge carriers at the surface of the gated material, as recently demonstrated in Ref.~\onlinecite{ZhangNanoLett2019}. We explicitly show that our ion-gated diamond films satisfy Eq.~\ref{eq:cap_series} in the inset to Fig.~\ref{figure:discussion}(a), and we fit the DOS-dependence of $C\ped{G}$ to Eq.~\ref{eq:cap_series} to determine $\alpha$ and $C\ped{H}$. From the best fit (black dashed line), we find $C\ped{H}\approx22\,\mu$F~cm\apex{-2}, a typical value for electrolyte-based gates~\cite{ZhangNanoLett2019, ChazalvielBook, PiattiPRM2019}, and $\alpha\approx0.015$. For comparison, the authors in Ref.~\cite{ZhangNanoLett2019} determined a much larger $\alpha\approx0.29\div0.47$ in ion-gated molybdenum and tungsten diselenides: This is consistent with their much larger values of $C\ped{G}\approx10\div40~\mu$F~cm\apex{-2} and comparable values of DOS $\approx2\div10\cdot10^{14}$~eV\apex{-1}cm\apex{-2} (depending on the number of filled bands). Since $\alpha = 1$ in the absence of cross-talk, this indicates that the influence of screening cross-talk between the ions and the field-induced charge carriers in the ion-gated diamond films is much stronger than in the ion-gated diselenides, likely due to subtle differences in electrostatic screening between the three materials. Although a quantitative, microscopical explanation for this difference will likely require a more detailed model than the one currently available~\cite{ZhangNanoLett2019}, it is interesting to note that an improved electrostatic screening due to a much thicker B-doped layer may help in explaining the origin of the much larger $C\ped{G}\simeq6.3~\mu$F~cm\apex{-2} observed in thick (300 nm) B-doped nanocrystalline diamond films~\cite{PiattiLTP2019, PiattiEPJ2019}. {\color{blue}We also stress that -- since all the ion-gated surfaces investigated in this work can be described by the same value of $\alpha$ irrespectively of the orientation and B content -- such a thickness-dependent screening cannot account for the orientation and B-doping dependencies of $C\ped{G}$ observed in our epitaxial films, which are instead fully described by the orientation-specific band filling in the (111) and (110) diamond surfaces. As a consequence, we expect the influence of any deviations of the B-doped layer thickness from the nominal value to be secondary in the samples investigated in our work.}

Gate-dependent electric transport measurements and \textit{ab initio} DFT calculations can also be combined to determine the doping-dependence of the Drude-like charge-carrier scattering lifetime $\tau$ from that of the mobility, disentangling it from the doping-dependencies of the DOS and the average effective mass in the system \cite{BrummePRB2016, PiattiApSuSc2017, Gonnelli2dMater2017}. Note that $\tau$ is the total scattering lifetime in the system, and thus for $T<0$ includes both elastic scattering terms (such as those coming from defects) and inelastic scattering terms (such as electron-phonon scattering). As originally demonstrated in Ref.~\onlinecite{BrummePRB2016}, this can be done by dividing the experimental value of $\mu$ by the mobility-to-scattering lifetime ratio, $\mu/\tau$, which can be computed directly from the \textit{ab initio} electronic bandstructure. Assuming that $\tau$ is independent on band index and momentum (constant relaxation time approximation), then-plane conductivity can be written as~\cite{BrummePRB2016}:
\begin{equation}
\sigma\ped{2D}=e^2 \tau \langle v\ped{F}^2\rangle \mathrm{DOS}(E\ped{F})
\end{equation}
where $\langle v\ped{F}^2\rangle$ is the average squared in-plane Fermi velocity, which we compute from the \textit{ab initio }bandstructure as follows. For each high-symmetry cut across the Brillouin zone $j$, we calculate the corresponding group velocity at the Fermi level for the $i$-th band $\varepsilon_i$, \mbox{$v_{j,i} = 1/\hbar\,\partial\varepsilon_i/\partial k_j|_{E=E\ped{F}}$}. The values of $v_{j,i}$ are then averaged over the different high-symmetry cuts and over the filled bands:
\begin{equation}
\langle v\ped{F}^2\rangle = \frac{\sum_{ij}g_i(v_{j,i})^2}{\sum_{ij}g_ig_j}
\label{eq:vf}
\end{equation}
where $g_i$ marks whether the $i$-th band is empty ($g_i=0$) or filled ($g_i=1$), and $g_j$ is the number of high-symmetry cuts [$g_j=2$ for the (111) surface and $g_j=3$ for the (110) surface]. For each value of $n\ped{2D}$, the maximum difference between $v_{j,i}$ measured along the different high-symmetry cuts is taken as the uncertainty. Thus, one obtains the $n\ped{2D}$-dependence of $\mu/\tau$ by:
\begin{equation}
\frac{\mu}{\tau} = \frac{1}{\tau}\frac{\sigma\ped{2D}}{en\ped{2D}} = \frac{e\langle v\ped{F}^2\rangle\mathrm{DOS}(E\ped{F})}{n\ped{2D}}
\label{eq:mu_tau}
\end{equation}
which we plot in Fig.~\ref{figure:discussion}(b) for both the (111) and (110) diamond surfaces. In both cases, $\mu/\tau$ decreases with increasing $n\ped{2D}$, indicating that part of the mobility decrease as a function of $n\ped{2D}$ is due to the doping-dependence of the bandstructure. In the (111) surface, since the same two bands are filled at any $n\ped{2D}\leq 2\cdot10^{14}$~h\apex{+}cm\apex{-2}, the decrease of $\mu/\tau$ with increasing hole doping is featureless and monotonic. Conversely, in the (110) surface a clear non-monotonic feature is present in the doping-dependence of $\mu/\tau$ in the range between $n\ped{2D}=3\cdot 10^{13}$ and $7\cdot 10^{13}$~h\apex{+}cm\apex{-2}, caused by the filling of the second topmost valence band. The doping-induced crossing of the band edge by the Fermi level should also result in corresponding ``kinks" in the doping-dependencies of $\sigma\ped{2D}$ and $\mu$~\cite{BrummePRB2016}, as was demonstrated in gated single- and few-layer graphene~\cite{YePNAS2011, PiattiApSuSc2017, Gonnelli2dMater2017} and transition-metal dichalcogenides~\cite{PiattiNanoLett2018, PiattiJPCM2019, ZhangNanoLett2019}. However, this doping range was not covered by values of $n\ped{2D}$ experimentally reached in our (110)-oriented films, preventing the experimental observation of these non-monotonicities in our films.

In Fig.~\ref{figure:discussion}(c) we plot the dependence of $\tau$ on $n\ped{2D}$ obtained by dividing the experimental values of $\mu$, shown in Fig.~\ref{figure:mobility}(b), by the corresponding values of $\mu/\tau$ obtained by linear interpolation of the data shown in Fig.~\ref{figure:discussion}(b) to the experimentally-measured values of $n\ped{2D}$. Broadly speaking, the scattering lifetime shows a similar behavior as that of the mobility, with a few notable differences. First, in the H-(110) surface at $n\ped{2D}\lesssim 1\cdot10^{13}$~h\apex{+}cm\apex{-2}, $\tau$ actually \textit{increases} with increasing $n\ped{2D}$. This indicates that, at very low doping, the improved screening due to the increase in carrier density is more important than both the increase in electron-phonon scattering~\cite{YamaguchiJPSJ2013, PiattiLTP2019, RezekAPL2006, KasuAPE2012} and the gate-induced increase in Coulomb scattering centers~\cite{PiattiEPJ2019, PiattiLTP2019, GallagherNatCommun2015, SaitoACSNano2015, Gonnelli2dMater2017, PiattiApSuSc2017, PiattiNanoLett2018, PiattiApSuSc2018mos2, OvchinnikovNatCommun2016, PiattiAPL2017, LuPNAS2018, PiattiPRM2019, PiattiJPCM2019}. At larger doping, $\tau$ again decreases with increasing $n\ped{2D}$ in both the H-(110) and H-(111) surfaces. In the B-doped surfaces, on the other hand, $\tau$ is basically doping-independent within the uncertainty range, suggesting that the scattering rate is for the most part caused by the introduction of the B dopants. As was observed for $\mu$, we find that the doping-dependence of $\tau$ in the H-(110) surface can be extrapolated smoothly to that in the B-(110) surface, whereas a sharp drop by more than one order of magnitude is observed going from the H-(111) to the B-(111) surface.

Finally, we combine the theoretical values of the chemical potential measured from the top of the valence band, $E\ped{v}-E\ped{F}$, with the values of $\tau$ to calculate the Ioffe-Regel parameter $x = (E\ped{v}-E\ped{F})\tau/\hbar$ and directly assess the metallicity in our films. We show the resulting $n\ped{2D}$-dependence of $x$ in Fig.~\ref{figure:discussion}(d), together with the Mott-Ioffe-Regel criterion $x=1$ which characterizes the insulator-to-metal transition (IMT) in disordered systems~\cite{IoffePS1960}. Once again, the behavior of the ion-gated H-terminated and B-doped surfaces is starkly different. In both the H-(111) and H-(110) surfaces, the additional doping provided via the ionic gate allows to tune the system from relatively deep in the insulating side of the IMT ($x<1$) to the quantum critical regime slighly on the metallic side of the IMT ($x\approx1$). In these conditions, we expect the conductivity of the system to decrease as a power law with decreasing temperature, but still remaining finite for $T\rightarrow0$. The introduction of B dopants, on the other hand, effectively ``freezes" the state of the two surfaces, preventing a further tuning via the ionic gate. The B-(110) surface becomes locked in the quantum critical regime, very close to the Mott-Ioffe-Regel criterion, effectively frustrating the full development of the IMT. The B-(111) surface is pushed back into the insulating side, resulting in a full-on re-entrant IMT instead, although in this case the ionic gate appears to be able to still slightly tune $x$ closer to the IMT, even if ineffectively. Overall, the effects of B doping are confirmed to be strongly dependent on the film orientation: The (110) surface is found to be very resilient against the introduction of extrinsic disorder caused by the B dopants, thus obtaining a main effect in the form of an additional hole doping. The (111) surface, on the other hand, while more effectively incorporating the B dopants in the crystal lattice, is also extremely sensitive to the additional extrinsic disorder, which is found to be the dominant effect in the tuning of its metallicity.

\section{Conclusions}\label{sec:conclusions}

In summary, we have investigated the impact of a hole co-doping approach, which combines ionic gating and standard B substitution, on the electric transport properties of H-terminated diamond films epitaxially grown on (111)- and (110)-oriented diamond substrates. By performing gate-dependent transport measurements at $T=240$~K, we have shown that the effect of B doping is strongly dependent on the crystal orientation. In the (111) surface, it strongly suppresses the charge-carrier mobility and leads to a small increase in the gate capacitance. In the (110) surface, it strongly increases the gate capacitance with a moderate reduction in mobility. In both surfaces, the maximum capacitance is much smaller than that of B-doped, thick nanocrystalline films, suggesting that the thickness of B-doped diamond plays a significant role in determining the gate capacitance. Our \textit{ab initio} DFT calculations of the electronic bandstructure indicate that the strongly orientation-dependent increase of the gate capacitance is due to the specific energy-dependence of the density of states in the two surfaces, resulting in very different doping dependencies of their quantum capacitances. Moreover, part of the reduction of mobility with increasing hole density stems from the doping-dependence of the bandstructure, as evidenced by the doping-dependence of the scattering lifetimes obtained combining the bandstructure calculations with the experimental values of the mobility. Additionally, our results indicate that, due to the orientation-dependent sensitivity towards extrinsic disorder introduced by the B dopants, our co-doping approach leads to a frustrated insulator-to-metal transition in the (110) surface, and to a fully re-entrant insulator-to-metal transition in the (111) surface. Further improvements towards a robust metallicity could be achieved by optimizing the crystal growth~\cite{ElHajjDRM2008} and the B doping process~\cite{OkazakiAPL2015} in order to avoid the presence of discontinuities in the doped surface layer, minimize the mobility degradation upon the introduction of B dopants, and increase the free carrier density. At the same time, the effectiveness of the ionic gate may be improved by introducing an ultrathin boron nitride spacer layer between the H-terminated surface and the electrolyte, as it would strongly increase the mobility of the H-terminated surfaces~\cite{SasamaAPLM2018, SasamaPRM2019} and help in increasing the quantum capacitance of the two surfaces~\cite{ZhangNanoLett2019}. A combination of all these methods will likely be required to further increase the total hole density per unit surface while retaining a large enough mobility for metallic behavior, key ingredients necessary to finally reach the field-induced superconductivity predicted for both surfaces~\cite{NakamuraPRB2013, SanoPRB2017, RomaninApSuSc2019, RomaninApSuSc2020}.

\section*{Acknowledgments}
We are very thankful to D. Romanin for his expertise and crucial help in performing the DFT calculations, and to C. Findler for operating the intrinsic CVD reactor. We also thank A. F. Morpurgo for fruitful scientific discussions concerning the quantum capacitance of ion-gated materials. This work was financially supported by the MIUR PRIN-2017 program (Grant No. 2017Z8TS5B - "Tuning and understanding
Quantum phases in 2D Materials - Quantum2D"). Computational resources were provided by HPC@polito (www.hpc.polito.it).


\begin{thebibliography}{99}
%
\bibitem{PanBook} L. S. Pan and D. R. Kania, \textit{Diamond: Electronic Properties and Applications} (Kluwer Academic Publishers, 1995).
%
\bibitem{EkimovNature2004} E. A. Ekimov, V. A. Sidorov, E. D. Bauer, N. N. Mel'nik, N. J. Curro, J. D. Thompson, and S. M. Stishov. Superconductivity in diamond. \textit{Nature} \textbf{428}, 542 (2004).
%
\bibitem{BustarretPRL2004} E. Bustarret, J. Ka{\v c}mar{\v c}ik, C. Marcenat, E. Gheeraert, C. Cytermann, J. Marcus, and T. Klein. Dependence of the Superconducting Transition Temperature on the Doping Level in Single-Crystalline Diamond Films. \textit{Phys. Rev. Lett.} \textbf{93}, 237005 (2004).
%
\bibitem{YokoyaNature2005} T. Yokoya, T. Nakamura, T. Matsushita, T. Muro, Y. Takano, M. Nagao, T. Takenouchi, H. Kawarada, and T. Oguchi. Origin of the metallic properties of heavily boron-doped superconducting diamond. \textit{Nature} \textbf{438}, 647 (2005).
%
\bibitem{TakanoAPL2004} Y. Takano, M. Nagao, I. Sakaguchi, M. Tachiki, and T. Hatano. Superconductivity in diamond thin films well above liquid helium temperature. \textit{Appl. Phys. Lett.} \textbf{85}, 2851 (2004).
%
\bibitem{IshizakaPRL2007} K. Ishizaka, R. Eguchi, S. Tsuda, T. Yokoya, A. Chainani, T. Kiss, T. Shimojima, T. Togashi, S. Watanabe, C.-T. Chen, C. Q. Zhang, Y. Takano, M. Nagao, I. Sakaguchi, T. Takenouchi, H. Kawarada, and S. Shin. Observation of a Superconducting Gap in Boron-Doped Diamond by Laser-Excited Photoemission Spectroscopy. \textit{Phys. Rev. Lett.} \textbf{98}, 047003 (2007).
%
\bibitem{BustarretPSSA2008} E. Bustarret. Superconducting diamond: an introduction. \textit{Phys. Status Solidi A} \textbf{205}, 997 (2008).
%
\bibitem{OkazakiAPL2015} H. Okazaki, T. Wakita, T. Muro, T. Nakamura, Y. Muraoka, T. Yokoya, S. Kurihara, H. Kawarada, T. Oguchi, and Y. Takano. Signature of high T\ped{c} above 25~K in high quality superconducting diamond. \textit{Appl. Phys. Lett.} \textbf{106}, 052601 (2015).
%
\bibitem{BoeriPRL2004} L. Boeri, J. Kortus, and O. K. Andersen. Three-Dimensional MgB\ped{2}-Type Superconductivity in Hole-Doped Diamond. \textit{Phys. Rev. Lett.} \textbf{93}, 237002 (2004).
%
\bibitem{LeePRL2004} K.-W. Lee and W. E. Pickett. Superconductivity in Boron-Doped Diamond. \textit{Phys. Rev. Lett.} \textbf{93}, 237003 (2004).
%
\bibitem{BoeriJPCS2006} L. Boeri, J. Kortus, and O. K. Andersen. Electron-phonon superconductivity in hole-doped diamond: A first-principles study. \textit{J. Phys. Chem. Solids} \textbf{67}, 552 (2006).
%
\bibitem{GiustinoPRL2007} F. Giustino, J. R. Yates, I. Souza, M. L. Cohen, and S. G. Louie. Electron-Phonon Interaction via Electronic and Lattice Wannier Functions: Superconductivity in Boron-Doped Diamond Reexamined. \textit{Phys. Rev. Lett.} \textbf{98}, 047005 (2007).
%
\bibitem{LandstrassAPL1989} M. I. Landstrass and K. V. Ravi. Resistivity of chemical vapor deposited diamond films. \textit{Appl. Phys. Lett.} \textbf{55}, 975 (1989).
%
\bibitem{MaierPRL2000} F. Maier, M. Riedel, B. Mantel, J. Ristein, and L. Ley. Origin of Surface Conductivity in Diamond. \textit{Phys. Rev. Lett.} \textbf{85}, 3472 (2000).
%
\bibitem{StrobelNature2004} P. Strobel, M. Riedel, J. Ristein and L. Ley. Surface transfer doping of diamond. \textit{Nature} \textbf{430}, 439 (2004).
%
\bibitem{EdmondsNanoLett2015} M. T. Edmonds, L. H. Willems van Beveren, O. Klochan, J. Cervenka, K. Ganesan, S. Prawer, L. Ley, A. R. Hamilton, and C. I. Pakes. Spin-Orbit Interaction in a Two-Dimensional Hole Gas at the Surface of Hydrogenated Diamond. \textit{Nano Lett.} \textbf{15}, 16-20 (2015).
%
\bibitem{NakamuraPRB2013} K. Nakamura, S. H. Rhim, A. Sugiyama, K. Sano, T. Akiyama, T. Ito, M. Weinert, and A. J. Freeman. Electric-field-driven hole carriers and superconductivity in diamond. \textit{Phys. Rev. B} \textbf{87}, 214506 (2013).
%
\bibitem{SanoPRB2017} K. Sano, T. Hattori, and K. Nakamura. Role of surface-bound hole states in electric-field-driven superconductivity at the (110)-surface of diamond. \textit{Phys. Rev. B} \textbf{96}, 155144 (2017).
%
\bibitem{RomaninApSuSc2019} D. Romanin, T. Sohier, D. Daghero, F. Mauri, R. S. Gonnelli, and M. Calandra. Electric field exfoliation and high-T\ped{c} superconductivity in field-effect hole-doped hydrogenated diamond (111). \textit{Appl. Surf. Sci.} \textbf{496}, 143709 (2019).
%
\bibitem{RomaninApSuSc2020} D. Romanin, G. A. Ummarino, and E. Piatti. Migdal-Eliashberg theory of multi-band high-temperature superconductivity in field-effect-doped (111) diamond. \textit{Appl. Surf. Sci.} \textbf{This issue}. Preprint available at arXiv:2002.11554.
%
{\color{blue}\bibitem{DankerlPRL2011} M. Dankerl, A. Lippert, S. Birner, E.U. St{\"u}tzel, M. Stutzmann, and J. A. Garrido. Hydrophobic Interaction and Charge Accumulation at the Diamond-Electrolyte Interface. \textit{Phys. Rev. Lett.} \textbf{106}, 196103 (2011).}
%
\bibitem{PiattiPRB2017} E. Piatti, D. Daghero, G. A. Ummarino, F. Laviano, J. R. Nair, R. Cristiano, A. Casaburi, C. Portesi, A. Sola, and R. S. Gonnelli. Control of bulk superconductivity in a BCS superconductor by surface charge doping via electrochemical gating. \textit{Phys. Rev. B} \textbf{95}, 140501(R) (2017).
%
\bibitem{FeteAPL2016} A. F{\^e}te, L. Rossi, A. Augieri, and C. Senatore. Ionic liquid gating of ultra-thin YBa\ped{2}Cu\ped{3}O\ped{7-x} films. \textit{Appl. Phys. Lett.} \textbf{109}, 192601 (2016).
%
\bibitem{PiattiApSuSc2018nbn} E. Piatti, D. Romanin, R. S. Gonnelli, and D. Daghero. Anomalous screening of an electrostatic field at the surface of niobium nitride. \textit{Appl. Surf. Sci.} \textbf{461}, 17 (2018).
%
\bibitem{UmmarinoPRB2017} G. A. Ummarino, E. Piatti, D. Daghero, R. S. Gonnelli, I. Yu. Sklyadneva, E. V. Chulkov, and R. Heid. Proximity Eliashberg theory of electrostatic field-effect-doping in superconducting films. \textit{Phys. Rev. B} \textbf{96}, 064509 (2017).
%
\bibitem{ValentinisPRB2017} D. Valentinis, S. Gariglio, A. F{\^e}te, J.-M. Triscone, C. Berthod, and D. van der Marel. Modulation of the superconducting critical temperature due to quantum confinement at the LaAlO\ped{3}/SrTiO\ped{3} interface. \textit{Phys. Rev. B} \textbf{96}, 094518 (2017).
%
\bibitem{PiattiLTP2019} E. Piatti, D. Romanin, D. Daghero, and R. S. Gonnelli. Two-dimensional hole transport in ion-gated diamond surfaces: A brief review. \textit{Low Temp. Phys.} \textbf{45}, 1143 (2019).
%
\bibitem{YamaguchiJPSJ2013} T. Yamaguchi, E. Watanabe, H. Osato, D. Tsuya, K. Deguchi, T. Watanabe, H. Takeya, Y. Takano, S. Kurihara, and H. Kawarada. Low-Temperature Transport Properties of Holes Introduced by Ionic Liquid Gating in Hydrogen-Terminated Diamond Surfaces. \textit{J. Phys. Soc. Jpn.} \textbf{82}, 074718 (2013).
%
\bibitem{TakahidePRB2014} Y. Takahide, H. Okazaki, K. Deguchi, S. Uji, H. Takeya, Y. Takano, H. Tsuboi, and H. Kawarada. Quantum oscillations of the two-dimensional hole gas at atomically flat diamond surfaces. \textit{Phys. Rev. B} \textbf{89}, 235304 (2014).
%
\bibitem{AkhgarNanoLett2016} G. Akhgar, O. Klochan, L. H. Willems van Beveren, M. T. Edmonds, F. Maier, B. J. Spencer, J. C. McCallum, L. Ley, A. R. Hamilton, and C. I. Pakes. Strong and Tunable Spin-Orbit Coupling in a Two-Dimensional Hole
Gas in Ionic-Liquid Gated Diamond Devices. \textit{Nano Lett.} \textbf{16}, 3768 (2016).
%
\bibitem{TakahidePRB2016} Y. Takahide, Y. Sasama, M. Tanaka, H. Takeya, Y. Takano, T. Kageura, and H. Kawarada. Spin-induced anomalous magnetoresistance at the (100) surface of hydrogen-terminated diamond. \textit{Phys. Rev. B} \textbf{94}, 161301(R) (2016).
%
\bibitem{AkhgarPRB2019} G. Akhgar, L. Ley, D. L. Creedon, A. Stacey, J. C. McCallum, A. R. Hamilton, and C. I Pakes. \textit{g}-factor and well-width fluctuations vs carrier density in the 2d hole accumulation layer of transfer-doped diamond. \textit{Phys. Rev. B} \textbf{99}, 035159 (2019).
%
\bibitem{DagheroPRL2012} D. Daghero, F. Paolucci, A. Sola, M. Tortello, G. A. Ummarino, M. Agosto, R. S. Gonnelli, J. R. Nair, and C. Gerbaldi. Large Conductance Modulation of Gold Thin Films by Huge Charge Injection via Electrochemical Gating. \textit{Phys. Rev. Lett.} \textbf{108}, 066807 (2012).
%
\bibitem{LiNature2016} L. J. Li, E. C. T. O'Farrel, K. P. Loh, G. Eda, B. {\"O}zyilmaz, and A. H. Castro Neto. Controlling many-body states by the electric-field effect in a two-dimensional material. \textit{Nature} \textbf{529}, 185 (2016).
%
\bibitem{XiPRL2016} X. Xi, H. Berger, L. Forr{\'o}, J. Shan, and K. F. Mak. Gate Tuning of Electronic Phase Transitions in Two-Dimensional NbSe\ped{2}. \textit{Phys. Rev. Lett.} \textbf{117}, 106801 (2016).
%
\bibitem{FujimotoPCCP2013} T. Fujimoto and K. Awaga. Electric-double-layer field-effect transistors with ionic liquids. \textit{Phys. Chem. Chem. Phys.} \textbf{15}, 8983 (2013).
%
\bibitem{UenoJPSJ2014} K. Ueno, H. Shimotani, H. Yuan, J. Ye, M. Kawasaki, and I. Iwasa. Field-Induced Superconductivity in Electric Double Layer Transistors. \textit{J. Phys. Soc. Jpn.} \textbf{83}, 032001 (2014).
%
\bibitem{PiattiEPJ2019} E. Piatti, F. Galanti, G. Pippione, A. Pasquarelli, and R. S. Gonnelli. Towards the insulator-to-metal transition in ion-gated nanocrystalline diamond films. \textit{Eur. Phys. J. Spec. Top.} \textbf{228}, 689 (2019).
%
\bibitem{PiattiJSNM2016} E. Piatti, A. Sola, D. Daghero, G. A. Ummarino, F. Laviano, J. R. Nair, C. Gerbaldi, R. Cristiano, A. Casaburi, and R. S. Gonnelli. Superconducting transition temperature modulation in NbN via EDL gating. \textit{J. Supercond. Novel Magn.} \textbf{29}, 587-591 (2016).
%
\bibitem{PiattiPRM2019} E. Piatti, T. Hatano, D. Daghero, F. Galanti, C. Gerbaldi, S. Guastella, C. Portesi, I. Nakamura, R. Fujimoto, K. Iida, H. Ikuta, and R. S. Gonnelli. Ambipolar suppression of superconductivity by ionic gating in optimally doped BaFe\ped{2}(As,P)\ped{2} ultrathin films. \textit{Phys. Rev. Materials} \textbf{3}, 044801 (2019).
%
\bibitem{TortelloApsusc2013} M. Tortello, A. Sola, K. Sharda, F. Paolucci, J. R. Nair, C. Gerbaldi, D. Daghero, and R. S. Gonnelli. Huge field-effect surface charge injection and conductance modulation in metallic thin films by electrochemical gating. \textit{Appl. Surf. Sci.} \textbf{269}, 17-22 (2013).
%
\bibitem{ScholtzBook} F. Scholtz. \textit{Electroanalytical Methods (2nd Edition).} Springer-Verlag, Berlin (2010).
%
\bibitem{ZhangNanoLett2019} H. Zhang, C. Berthod, H. Berger, T. Giamarchi, and A. F. Morpurgo. Band Filling and Cross Quantum Capacitance in Ion-Gated Semiconducting Transition Metal Dichalcogenide Monolayers. \textit{Nano Lett.} \textbf{19}, 8836-8845 (2019).
%
\bibitem{RezekAPL2006} B. Rezek, H. Watanabe, and C. E. Nebel. High carrier mobility on hydrogen terminated $\langle100\rangle$ diamond surfaces. \textit{Appl. Phys. Lett.} \textbf{88}, 042110 (2006). 
%
\bibitem{KasuAPE2012} M. Kasu, H. Sato, and K. Hirama. Thermal Stabilization of Hole Channel on H-Terminated Diamond Surface by Using Atomic-Layer-Deposited Al\ped{2}O\ped{3} Overlayer and its Electric Properties. \textit{Appl. Phys. Espress} \textbf{5}, 025701 (2012).
%
%
%
%
%
%
%
\bibitem{ElHajjDRM2008} H. El-Hajj, A. Denisenko, A. Bergmaier, G. Dollinger, M. Kubovic, and E. Kohn. Characteristics of boron $\delta$-doped diamond for electronic applications. \textit{Diamond Relat. Mater.} \textbf{17}, 409-414 (2008).
%
\bibitem{ScharpfPSSA2013} J. Scharpf, A. Denisenko, C. I. Pakes, S. Rubanov, A. Bergmaier, G. Dollinger, C. Pietzka, and E. Kohn. Transport behaviour of boron delta-doped diamond. \textit{Phys. Status Solidi A} \textbf{210}, 2028 (2013).
%
\bibitem{UshizawaDRM1998} K. Ushizawa, K. Watanabe, T. Ando, I. Sakaguchi, M. Nishitani-Gamo, Y. Sato, and H. Kanda. Boron concentration dependence of Raman spectra on {100} and {111} facets of B-doped CVD diamond. \textit{Diamond Relat. Mater.} \textbf{7}, 1719 (1998).
%
%
%
%
%
%
%
%
%
%
%
\bibitem{QEspresso} P. Giannozzi et al. Advanced capabilities for materials modelling with Quantum ESPRESSO. \textit{J. Phys. Condens. Matter} \textbf{29}, 465901 (2017).
%
\bibitem{PerdewPRL1996} J. P. Perdew, K. Burke, and M. Ernzerhof. Generalized Gradient Approximation Made Simple. \textit{Phys. Rev. Lett.} \textbf{77}, 3865 (1996).
%
\bibitem{BlochlPRB1994} P. E. Bl{\"o}chl. Projector augmented-wave method. \textit{Phys. Rev. B} \textbf{50}, 17953 (1994).
%
\bibitem{HamannPRB2013} D. R. Hamann. Optimized norm-conserving Vanderbilt pseudopotentials. \textit{Phys. Rev. B} \textbf{88}, 085117 (2013).
%
\bibitem{SSSP} K. Lejaeghere et al. Reproducibility in density functional theory calculations of solids. \textit{Science} \textbf{351}, 1415 (2016). G. Prandini, A. Marrazzo, I. E. Castelli, N. Mounet, and N. Marzari. Precision and efficiency in solid-state pseudopotential calculations. \textit{npj Comput. Mater.} \textbf{4}, 72 (2018). WEB: http://materialscloud.org/sssp.
%
\bibitem{MonkhorstPRB1976} H. J. Monkhorst and J. D. Pack. Special points for Brillouin-zone integrations. \textit{Phys. Rev. B} \textbf{13}, 5188 (1976).
%
\bibitem{SohierPRB2017} T. Sohier, M. Calandra, and F. Mauri. Density functional perturbation theory for gated two-dimensional heterostructures: Theoretical developments and application to flexural phonons in graphene. \textit{Phys. Rev. B} \textbf{96}, 075448 (2017).
%
%
%
%
%
%
%
%
%
%
%
%
%
%
%
%
%
%
%
%
%
%
%
%
%
%
%
%
\bibitem{LuryiAPL1988} S. Luryi. Quantum capacitance devices. \textit{Appl. Phys. Lett.} \textbf{52} 501-503 (1988).
%
\bibitem{EisensteinPRL1992} J. P. Eisenstein, L. N. Pfeiffer, and K. W. West. Negative compressibility of interacting two-dimensional electron and quasi-particle gases. \textit{Phys. Rev. Lett.} \textbf{68}, 674-677 (1992).
%
\bibitem{GiulianiBook2005} G. Giuliani and G. Vignale. \textit{Quantum Theory of the Electron Liquid}. Cambridge University Press (Cambridge, 2005).
%
\bibitem{IlaniNatPhys2006} S. Ilani, L. A. K. Donev, M. Kindermann, and P. L. McEuen. Measurement of the quantum capacitance of interacting electrons in carbon nanotubes. \textit{Nat. Phys.} \textbf{2}, 687-691 (2006).
%
\bibitem{YuPNAS2013} G. L. Yu, R. Jalil, B. Belle, A. S. Mayorov, P. Blake, F. Schedin, S. V. Morozov, L. A. Ponomarenko, F. Chiappini, S. Wiedmann, U. Zeitler, M. I. Katsnelson, A. K. Geim, K. S. Novoselov, and D. C. Elias. Interaction phenomena in graphene seen through quantum capacitance. \textit{Proc. Natl. Acad. Sci. USA} \textbf{110}, 3282-3286 (2013).
%
\bibitem{UesugiSciRep2013} E. Uesugi, H. Goto, R. Eguchi, A. Fujiwara, and Y. Kubozono. Electric double-layer capacitance between an ionic liquid and few-layer graphene. \textit{Sci. Rep.} \textbf{3}, 1595 (2013).
%
\bibitem{ChuSciRep2014} L. Chu, H. Schmidt, J. Pu, S. Wang, B. {\"O}zylmaz, T. Takenobu, and G. Eda. Charge transport in ion-gated mono-, bi- and trilayer MoS\ped{2} field effect transistors. \textit{Sci. Rep.} \textbf{4}, 7293 (2014).
%
\bibitem{Gonnelli2dMater2017} R. S. Gonnelli, E. Piatti, A. Sola, M. Tortello, F. Dolcini, S. Galasso, J. R. Nair, C. Gerbaldi, E. Cappelluti, M. Bruna, and A. C. Ferrari. Weak localization in electric-double-layer gated few-layer graphene. \textit{2D Mater.} \textbf{4}, 035006 (2017).
%
\bibitem{PiattiApSuSc2017} E. Piatti, S. Galasso, M. Tortello, J. R. Nair, C. Gerbaldi, M. Bruna, S. Borini, D. Daghero, and R. S. Gonnelli. Carrier mobility and scattering lifetime in electric double-layer gated few-layer graphene. \textit{Appl. Surf. Sci.} \textbf{395}, 37 (2017).
%
\bibitem{YePNAS2011} J. Ye, M. F. Craciun, M. Koshino, S. Russo, S. Inoue, H. Yuan, H. Shimotani, A. F. Morpurgo, and Y. Iwasa. Accessing the transport properties of graphene and its multilayers at high carrier density. \textit{Proc. Natl. Acad. Sci. USA} \textbf{108}, 13002 (2011).
%
\bibitem{GonnelliSciRep2015} R. S. Gonnelli, F. Paolucci, E. Piatti, K. Sharda, A. Sola, M. Tortello, J. R. Nair, C. Gerbaldi, M. Bruna, and S. Borini. Temperature Dependence of Electric Transport in Few-layer Graphene under Large Charge Doping Induced by Electrochemical Gating. \textit{Sci. Rep.} \textbf{5}, 9554 (2015).
%
\bibitem{LeNS2019} S. T. Le, N. B. Guros, R. C. Bruce, A. Cardone, N. D. Amin, S. Zhang, J. B. Klauda, H. C. Pant, C. A. Richter, and A. Balijepalli. Quantum capacitance-limited MoS\ped{2} biosensors enable remote label-free enzyme measurements. \textit{Nanoscale} \textbf{11}, 15622-15632 (2019).
%
\bibitem{ChazalvielBook} J.-N. Chazalviel, \textit{Coulomb Screening by Mobile Charges: Applications to Materials Science, Chemistry, and Biology}
(Springer, New York, 1999).
%
\bibitem{BrummePRB2016} T. Brumme, M. Calandra, and F. Mauri. Determination of scattering time and of valley occupation in transition-metal dichalcogenides doped by field effect. \textit{Phys. Rev. B} \textbf{93}, 081407(R) (2016).
%
\bibitem{IoffePS1960} A. F. Ioffe and A. R. Regel. Non-crystalline, amorphous and liquid electronic semiconductors. \textit{Prog. Semicond.} \textbf{4}, 237 (1960).
%
\bibitem{SasamaAPLM2018} Y. Sasama, K. Komatsu, S. Moriyama, M. Imura, T. Teraji, K. Watanabe, T. Taniguchi, T. Uchihashi, and Y. Takahide. High-mobility diamond field effect transistor with a monocrystalline h-BN gate dielectric. \textit{APL Mater.} \textbf{6}, 111105 (2018).
%
\bibitem{SasamaPRM2019} Y. Sasama, K. Komatsu, S. Moriyama, M. Imura, S. Sogiura, T. Terashima, S. Uji, K. Watanabe, T. Taniguchi, T. Uchihashi, and Y. Takahide. Quantum oscillations in diamond field-effect transistors with a h-BN gate dielectric. \textit{Phys. Rev. Mater.} \textbf{3}, 121601(R) (2019).
%
%
%
%
%
%
%
%
%
%
%
%
%
%
%
%
%
%
%
\bibitem{GallagherNatCommun2015} P. Gallagher, M. Lee, T. A. Petach, S. W. Stanwyck, J.R. Williams, K. Watanabe, T. Taniguchi, and D. Goldhaber-Gordon. A high-mobility electronic system at an electrolyte-gated oxide surface. \textit{Nat. Commun.} \textbf{6}, 6437 (2015).
%
\bibitem{SaitoACSNano2015} Y. Saito and Y. Iwasa. Ambipolar Insulator-to-Metal Transition in Black Phosphorus by Ionic-Liquid Gating. \textit{ACS Nano} \textbf{9}, 3192 (2015).
%
\bibitem{PiattiNanoLett2018} E. Piatti, D. De Fazio, D. Daghero, S. R. Tamalampudi, D. Yoon, A. C. Ferrari, and R. S. Gonnelli. Multi-Valley Supercondutivity in Ion-Gated MoS\ped{2} Layers. \textit{Nano Lett.} \textbf{18}, 4821 (2018).
%
\bibitem{PiattiJPCM2019} E. Piatti, D. Romanin, and R. S. Gonnelli. Mapping multi-valley Lifshitz transitions induced by field-effect doping in strained MoS\ped{2} nanolayers. \textit{J. Phys. Condens. Matter} \textbf{31}, 114002 (2019).
%
\bibitem{PiattiApSuSc2018mos2} E. Piatti, Q. Chen, M. Tortello, J. Ye, and R. S. Gonnelli. Possible charge-density-wave signatures in the anomalous resistivity of Li-intercalated multilayer MoS\ped{2}. \textit{Appl. Surf. Sci.} \textbf{461}, 269 (2018).
%
\bibitem{OvchinnikovNatCommun2016} D. Ovchinnikov, F. Gargiulo, A. Allain, D. J. Pasquier, D. Dumcenco, C.-H. Ho, O. V. Yazyev, and A. Kis. Disorder engineering and conductivity dome in ReS\ped{2} with electrolyte gating. \textit{Nat. Commun.} \textbf{7}, 12391 (2016)
%
\bibitem{PiattiAPL2017} E. Piatti, Q. Chen, and J. Ye. Strong dopant dependence of electric transport in ion-gated MoS\ped{2}. \textit{Appl. Phys. Lett.} \textbf{111}, 013106 (2017).
%
\bibitem{LuPNAS2018} J. Lu, O. Zheliuk, Q. Chen, I. Leermakers, N. E. Hussey, U. Zeitler, and J. Ye. Full superconducting dome of strong Ising protection in gated monolayer WS\ped{2}. \textit{Proc. Natl. Acad. Sci. USA} \textbf{115}, 3551 (2018).
%
\end{thebibliography}
\end{document}